# RASP FOR LSASS: PREVENTING MIMIKATZ-RELATED ATTACKS

Anna Revazova
Bachelor of Information Security, MEPhI
Moscow, Russia
aanrevazova@gmail.com

Igor Korkin, PhD
Security Researcher
Moscow, Russia
igor.korkin@gmail.com

## ABSTRACT

The Windows authentication system is implemented using the LSA (Local Security Authority) system and its process lsass.exe. This structure has a number of vulnerabilities, which makes it extremely attractive to intruders. Using known vulnerabilities from the CVE database or special mimikatz-type software, an attacker is able to obtain the user's password-address information. This paper will consider some ways to prevent attacks on local authentication subsystem as well as show some practical results of some methods, including those formulated earlies, and consider their applicability in real life.

## 1 INTRODUCTION

Despite the growing popularity of other operating systems, nowdays the Windows operating system is one of the most popular and commonly used for PCs both for personal use and for developers. The process of local user authentication is implemented in the authentication service of the local security system. The Local Security System Authentication Service (LSASS) is a part of the Windows operating system responsible for authorization of local users of a separate computer. The service is critical, since without it, logging in for local users (not registered in the domain) is impossible in principle. A very simple structure of Windows authentication is shown on Figure 1.

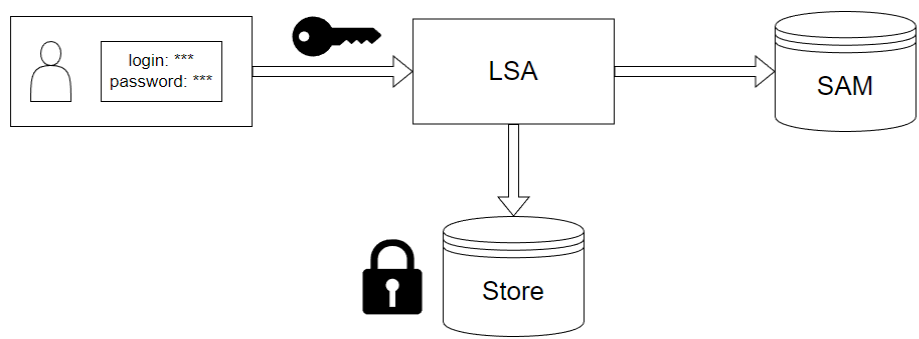

Figure 1 — Simple scheme of authentication

The process checks the authorization data, and for the successful authorization, the service sets a flag about the possibility of logging in. If authorization was started by the user, the user shell startup flag is also set. If authorization was initialized by a service or application, this application is granted the rights of this user.

When this service becomes infected with viruses or when intruder gets full access to this service, the system obeys the attacker, who can get full rights to access the target computer. Therefore, the encryption method and the method of transferring authorization data between components are not documented. In addition, the password is not transmitted in pure form, but in the form of a hash, which is compared with the hash of the real password.

For further work, we should consider the local authentication scheme in a little more detail.

First of all, let's start with local authentication, when a user wants to log in directly to a workstation that is not part of the domain. What happens after the user has entered their username and password? Immediately after that, the entered data is transmitted to the local security subsystem (LSA), which immediately converts the password into a hash (hashing is a one-way cryptographic transformation that makes it impossible to restore the original sequence). In plain text, the password is not stored anywhere in the system and does not appear, the user is the only one who knows it.

Then the LSA service contacts the Security Account Manager (SAM) and tells him the username. The dispatcher accesses the SAM database and extracts from there the password hash of the specified user generated when creating an account (or in the process of changing the password). Then the LSA compares the hash of the entered password and the hash from the SAM database, if they match, authentication is considered successful, and the hash of the entered password is placed in the storage of the LSA service and remains there until the end of the user session.

At the moment, one of the most well—known and easy-to-use software for dumping user credentials is mimikatz. Mimikatz is an open-source application that allows users to view and save credentials and other password-address information.

To use it, it is enough to install the archive on the official website or compile it yourself from the source code. In this paper, the second option was used.

The standard antivirus installed on the user's machine is able to detect mimikatz even at the download stage, but this protection can be bypassed with code obfuscation. To do this, it is enough to rename the downloaded archive and change some words in source code of mimikatz.

## 2 LITERATURE REVIEW

Microsoft researchers have identified the most popular data collection methods [3]. These methods are presented in Table 1.

Table 1 — Credential technique

| Attack techniques and tools | Description |
|---|---|
| Comsvc.dll (and its "MiniDump" export) loaded by rundll32.exe | Rundll allows to run the built-in Windows library comsvcs.dll, which exports a function called MiniDump. After calling this function, an attacker can submit the LSASS process PID as a parameter and create its memory dump |
| Mimikatz (and its modified variants) | An open-source application that allows users to view and save authentication credentials |
| Procdump.exe (with -ma command line option) | A tool included in the Windows Sysinternals package that allows to create process dumps which could be necessary to study some problems |
| Taskmgr.exe | The task manager allows to dump a process if it is running in privileged mode |

In this paper, we will focus in more detail on attacks using mimikatz.

Initially, it was planned to delete password-address information directly from the process memory lsass.exe, as well as residual information on the carrier. But mimikatz does not work with residual information, but directly with the process itself or its dump, so overwriting the residual information does not help and the user's authentication data in open form can still be obtained by a third party.

On the other hand, if we try to manually affect the process lsass.exe in order to change the user's data, the system will not allow the user to log in after a reboot, because when we logging in the entered parameters are checked with data that stored in the process lsass.exe, namely in the SAM subsystem. If we manually interact with this process, then there is a risk of breaking the system, after which only a complete reinstallation of Windows OS will be able to return it to its working state. Most likely, if you try to log in again, you will get a black screen with a white cursor on it.

At the beginning of its work, mimikatz connects to samsrv.dll, then uses the LsaOpenPolicy() function to request police descriptor. In Figure 2 there is part of mimikatz source-code [1], that shows the usage of LsaOpenPolicy() function. In Figure 3a there is an example of debugging of mimikatz so we find this function.

```
if(NT_SUCCESS(LsaOpenPolicy(szSystem ? &uSystem : NULL, &oaLsa, POLICY_VIEW_LOCAL_INFORMATION, &hLSA)))
{
    status = LsaQueryInformationPolicy(hLSA, PolicyDnsDomainInformation, (PVOID *) &pDomainInfo);
    if(NT_SUCCESS(status))
    {
```

Figure 2 — LsaOpenPolicy() in source code

The arguments of this function contain PolicyDnsDomainInformation, where we request information about the domain. After that we should pay attention to the following calls: SamConnect() and SamOpenDomain.

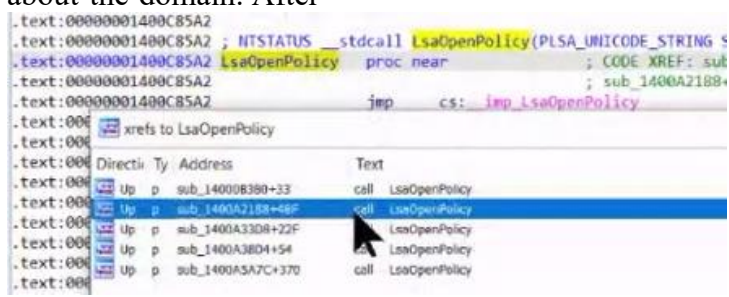

Figure 3a — LsaOpenPolicy() in IDA

As it mentioned before, mimikatz works with the SAM process, from where it extracts all the information of the WDigest, just in these functions it is performed. The SAM database server is implemented in the library samsrv.dll, and the functions are implemented in samlib.dll. As it can be seen from Figure 3b, mimikatz successfully connects with these libraries.

```
7ff72e530000 ce9d359b Nov 05 14:29:31 2079 C:\Windows\system32\lsass.exe
7fff44000000 57b668f2 Aug 19 05:03:30 2016 C:\Windows\SYSTEM32\ntdll.dll
7fff43090000 afec8296 Jul 13 03:38:46 2063 C:\Windows\System32\KERNEL32.DLL
7fff414e0000 b42fa627 Oct 17 19:13:27 2065 C:\Windows\System32\KERNELBASE.dll
7fff41e90000 c1879a9e Nov 20 20:47:10 2072 C:\Windows\System32\RPCRT4.dll
7fff40f30000 94f3cacb Mar 10 19:05:31 2049 C:\Windows\system32\lsasrv.dll
7fff41860000 00e78ce9 Jun 25 18:14:49 1970 C:\Windows\System32\ucrtbase.dll
7fff41aa0000 1fb7fd57 Nov 12 06:53:59 1986 C:\Windows\System32\msvcp_win.dll
7fff40ef0000 598cef6e Aug 11 02:42:38 2017 C:\Windows\system32\LSAADT.dll
7fff42e60000 62cba37a Jul 11 07:13:46 2022 C:\Windows\System32\sechost.dll
7fff40e00000 e804bd3e May 08 15:38:22 2093 C:\Windows\SYSTEM32\samsrv.dll
7fff41d20000 788edb3d Feb 04 04:09:17 2034 C:\Windows\System32\CRYPT32.dll
7fff40db0000 54fe428f Mar 10 04:02:07 2015 C:\Windows\system32\bcrypt.dll
7fff40d80000 4f134e55 Jan 16 01:08:21 2012 C:\Windows\system32\ncrypt.dll
7fff40d40000 7803c2df Oct 21 16:00:15 2033 C:\Windows\system32\NTASN1.dll
7fff40cf0000 27779f45 Dec 25 21:49:41 1990 C:\Windows\system32\Wldp.dll
```

Figure 3b — samsrv.dll in windbg

After that, let's pay attention to the name argument, from which we conclude that this function pulls out the domain name of the user.

We go further through the code and see a very interesting point: used samsrv.dll. Thus, our assumption that mimikatz uses one of the libraries of those group turned out to be correct, and inside the process lsass.exe a specific library is being launched

At the moment, among the most popular ways to combat data dumping from LSASS are the following recommendations presented by Microsoft [3]:

1. Use Windows defender. This method does not show itself very well, because it has been verified that this protection can be bypassed by obfuscating the source code and deleting/replacing keywords like "mimikatz" in it, thanks to such obfuscation, Microsoft Defender is not always able to successfully detect an attack. But some recent researches shows that Microsoft improves their protecting system and Windows Defender can work better.

2. Using PPL. This setting is disabled by default in Windows 10.

3. Enable Credential Guard in Windows Defender.

4. Enable Restricted Administrator mode for Remote Desktop Protocol (RDP).

5. Disable "UseLogonCredential" in WDigest. This method is not enough, since it can only stop attacks on passwords in plain text (for example, using the sekurlsa:logonpasswords command), but NTLM hashes can still be successfully obtained.

There are some extra methods to prevent mimikatz-based attacks made by independent researchers, that also have different level of applicability in real live, shown in table 2.

Table 2 — Extra methods to prevent mimikatz-based attacks

| Prevention methods | Description |
| --- | --- |
| Debug Privelege | Mimikatz requires this privilege as it interacts with processes such as LSASS. It is therefore important to set this privilege only to the specific group of people that will need this permission and remove it from the Local Administrators. |
| Disable LM and NTLM | These protocols are outdated, so you can refuse to use them. However, Windows OS has the property of backward compatibility, so if you disable these protocols, some of the functions will stop working. Moreover, mimikatz of the latest versions also successfully interacts with the kerberos protocol. |
| Protected Users | When using the functional level of the Windows Server 2012 R2 domain, it is possible to use a special protected group Protected Users to protect privileged users. In particular, the protection of these accounts from compromise is carried out due to the fact that members of this group can only log in via Kerberos. |

Also, some researchers introduce the concept of so-called honey tokens, which are capable of knocking an intruder off the right path [4] as one of the methods of detecting and preventing attacks. This method will be discussed in more detail in Chapter 4.

As an example, we will demonstrate the simplest attack scenario with the following starting condition: we have access to the victim's computer just to show how mimikatz works with simple scheme in Figure 4.

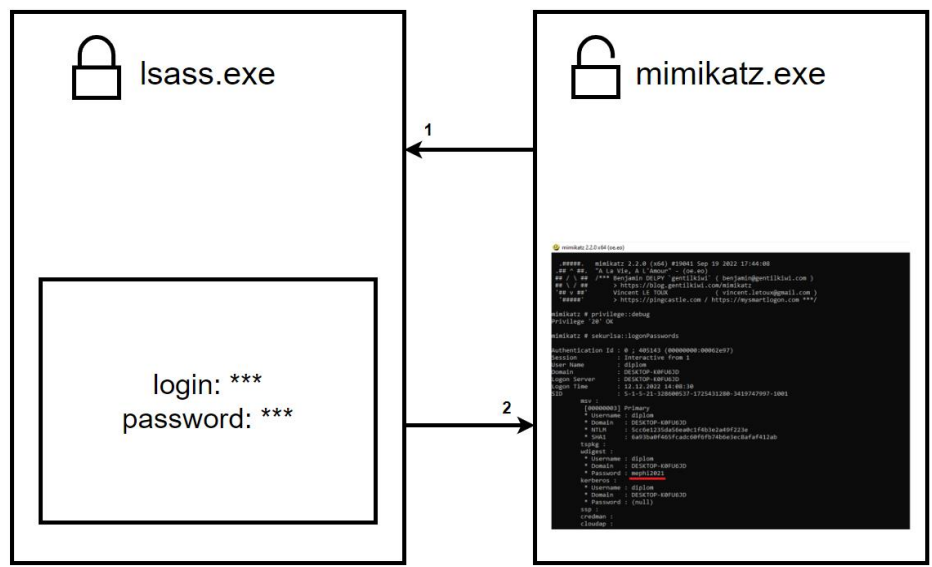

Figure 4 — Work of mimikatz.exe

To extract passwords in plain text, it is enough to use the following command:

sekurlsa::logonPasswords;

In order to use this command, it needs to increase privileges. To do this, also in the mimikatz console, the following command is executed:

privilege::debug;

As a result, you can get the user's passwords in plain text. The result is shown in Figure 5.

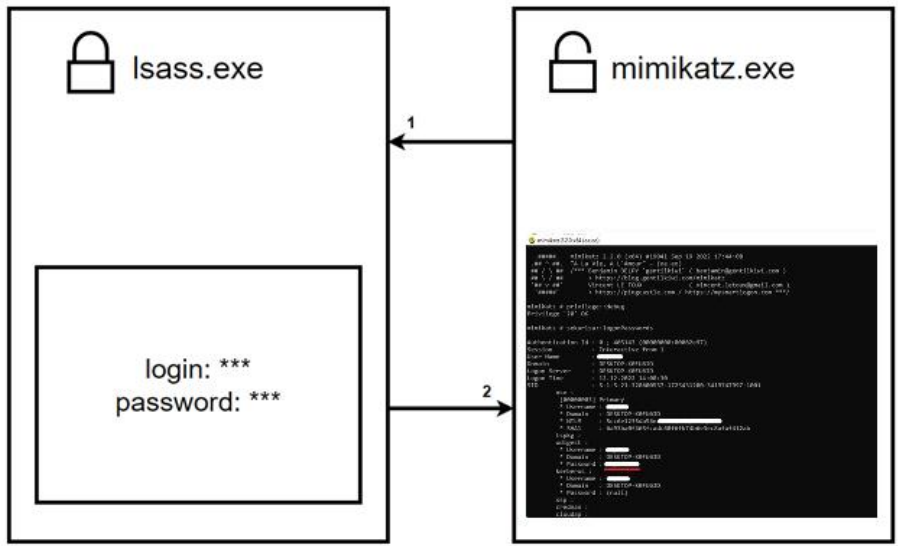

Figure 5 — extracting the password in plain text

There is a stronger command, thanks to which NTLM hashes are extracted, by performing the injection: lsadump::lsa /inject.

Figure 6 — Extracting the NTLM hash

NTLM is a subsystem that allows you to use the structure of "one user — one password" and "single sign—on", while you do not need to have knowledge of the internal structure. An NTLM hash is a hashed user password that was previously set and entered by the user. An example of extracting this hash is shown in Figure 6.

Thus, the user's password and address information become known.

## 3 PREVENTION TECHNIQUES

To combat mimikatz, we can offer 6 ways of varying degrees of complexity of implementation.

The easiest way is to introduce a *special user* into the process, which will distract the attacker's attention. Windows OS has several different functions that, in theory, can be used for these purposes according to their documentstion: CreateProcessAsUserA, CreateProcessWithLogonW,

CreateProcessAsUserW. As it turned out, the CreateProcessWithLogonW function is very easy to use for this purpose and it does not require any additional intervention. However, as it turned out, this method, which actually implements honey tokens, turned out to be extremely ineffective at a deeper level of attacks, as will be reported in the next chapter.

The second, rather time-consuming way is *scanning all processes* — to monitor all loaded dlls, followed by intercepting the call of certain functions (in particular LsaOpenPolicy()). In this case, it is required to check all possible programs for calling these functions, which is extremely inefficient. The general scheme of this application is shown on the Figure 7.

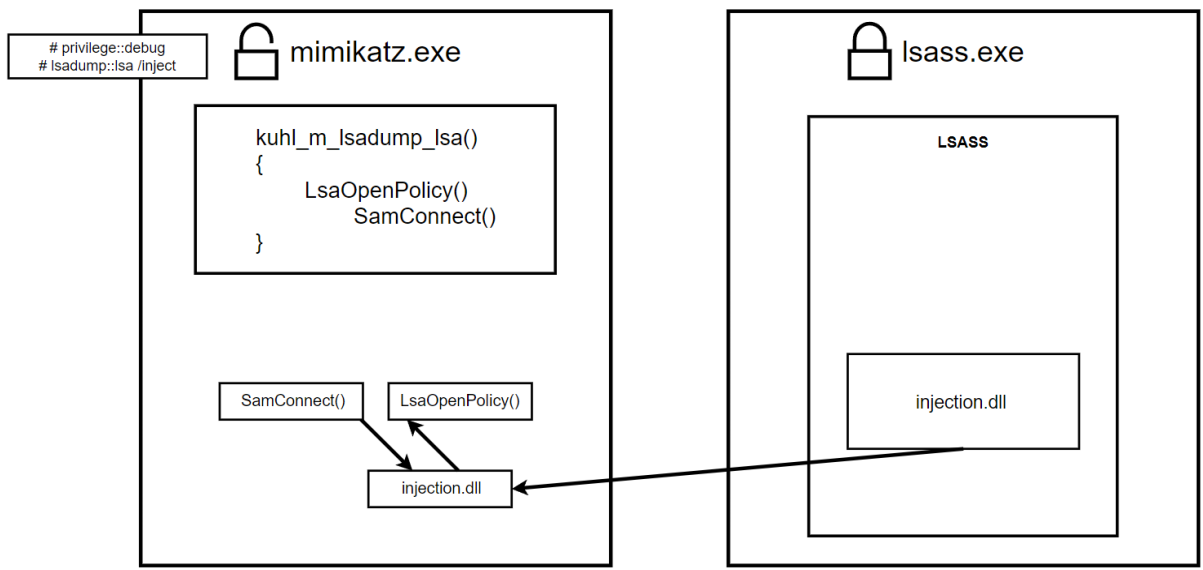

Figure 7 — General scheme

In a third way, we can suggest using *injection into LSASS*. We are making injection into the process lsass.exe and we prescribe the condition that if functions from libraries are called, firstly from samlib.dll and samsrv.dll the lock will be triggered. This method may be difficult to implement due to special process security requirements for lsass.exe. However, it can be quite effective.

Thanks to the work [5], it was found out that it is possible to block the creation of so-called ALPC connections of the SamConnect() function with the LSASS process and make *ALPC blocking*. However, this method will require some caution, since it is necessary to configure the return of the parameters to the original state, otherwise, after log-out from the system, we risk not entering it again.

The fifth suggestion is a kind of modification of the fourth — *injection to prevent ALPC connection*. It suggests injecting into LSASS in order to catch ALPC connections. This method has not been evaluated from a theoretical point of view, so its implementation is being questioned.

The last way is to work directly with variables samsrv.dll, namely *samsrv!SampServiceState*. This method has also not been analyzed.

Table 3 provides a complete analysis of mentioned methods that can be offered as a means of protection.

Table 3 — League table

| Method | Advantages | Disadvantages |
| --- | --- | --- |
| Special user — user that was written into LSASS | Simplicity of implementation, does not require the use of a large number of resources | It works only in the case of the simplest attacks, in fact, it does not completely hide information |
| Scanning all processes — scan all processes that can be used by mimikatz, especially mimikatz.exe itself | Guaranteed to find traces of using mimikatz | It is required to scan all applications running by the system, which requires significant resources |
| Injection into LSASS — injection into LSASS to check what functions are called | The golden mean in terms of quality and time | Requires the introduction of LSASS into the process which must be carried out with caution |
| ALPC blocking — block new ALPC connections between SamConnect() and LSASS | Requires rewriting three parameters for implementation | If used carelessly, there is a risk of not being able to log in to your account later |

## 4 PRACTICAL RESULTS

As previously reported, in this section we will consider in more detail the *special user* method with the introduction of an additional user who will distract attention to himself and thanks to which it will be possible to detect an intrusion. Among all the Windows functions, CreateProcessWithLogonW turned out to be the simplest, its signature is shown in Figure 8.

```
BOOL CreateProcessWithLogonW(
  [in]                LPCWSTR              lpUsername,
  [in, optional]      LPCWSTR              lpDomain,
  [in]                LPCWSTR              lpPassword,
  [in]                DWORD                dwLogonFlags,
  [in, optional]      LPCWSTR              lpApplicationName,
  [in, out, optional] LPWSTR               lpCommandLine,
  [in]                DWORD                dwCreationFlags,
  [in, optional]      LPVOID               lpEnvironment,
  [in, optional]      LPCWSTR              lpCurrentDirectory,
  [in]                LPSTARTUPINFOW       lpStartupInfo,
  [out]               LPPROCESS_INFORMATION lpProcessInformation
);
```

Figure 8 — Signature of CreateProcessWithLogonW

In fact, this function starts the process under the name of another user whose parameters are passed in this function. When mimikatz tries to access the user's passwords, it fails because the credentials of the embedded user fall out of the foreground as the highest priority. The general scheme of work is shown in Figure 9.

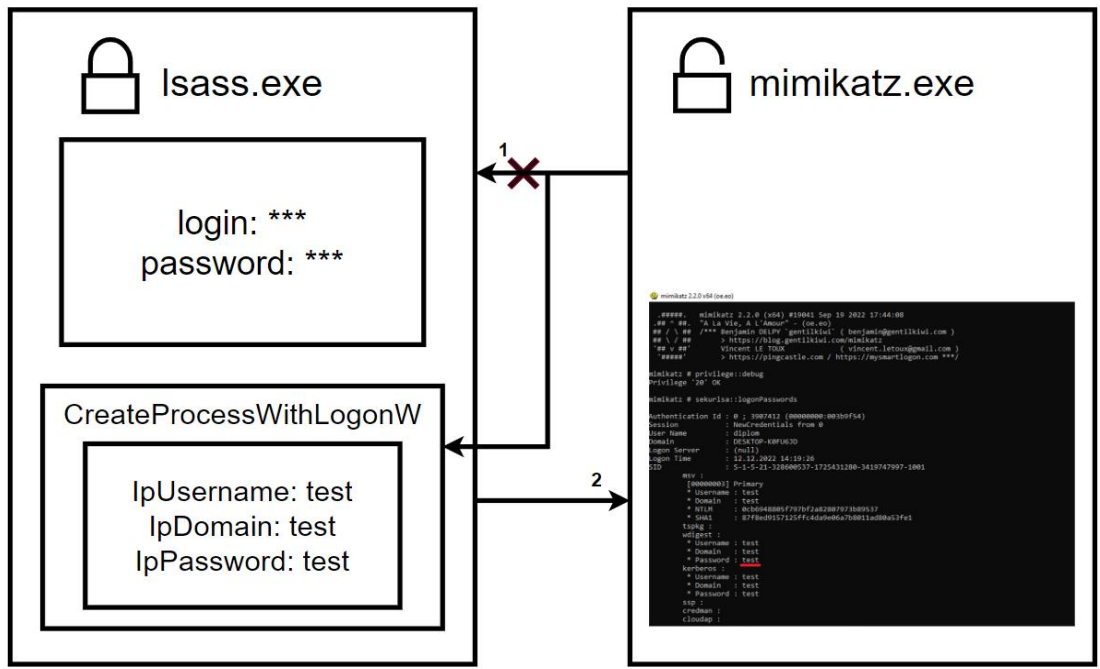

Figure 9 — General scheme

This function creates a new process on behalf of the user, whose parameters are specified in the Ip Username and IpPassword fields. To do this, you must first create an agent — a process that performs empty work, it will be launched on behalf of another user with the following parameters specified in Figure 9:

Login: test;

Domain: test;

Password: test.

After running mimikatz, you can make sure that this system is working, it turns out that mimikatz finds the password-address information of an artificially embedded user.

This injection can work well in a corporate system, it is enough to use such a fake user, which is known only to the system administrator, and then monitor the interaction with this user, which will mean the presence of an intruder on the network. However, this system has a drawback: along with the embedded user, a process that works in the system on an ongoing basis.

Let's check the operation of this system when performing an injection using mimikatz with next command:

lsadump::lsa /inject;

The result is shown on Figure 10.

Figure 10 — Result of lsadump::lsa /inject

Figure 10 shows that this system is able to "deceive" mimikatz with embedded data by lifting up the test user, but firstly is not able to close NTLM hashes and secondly, we can see the main user but in lower positions. It proves that this method doesn't work properly.

Thus, the method, which is often positioned as completely hiding password information that is valuable to an attacker, actually does not hide information as well as we would like. However, this method works well for detecting intrusions by additional logging.

Since the data that has been placed on the top, there is a possibility that an attacker is using it to attempt to log in. In this way, we create logging of login events under this account, after which it is enough to periodically view the file in which these logs are displayed or create a notification to this user logging-in.

The structure of the second method was shown in the figure 7.

When Mimikatz is launched, various dynamic libraries are connected. The LsaOpenPolicy function is accessed later, which gives us a very small amount of time advantage, after which information is requested from the plug-in libraries.

In this system, in turn, a manually written dynamic library is being implemented with specified call substitution actions, while the protection system will not only scan the applications being launched, but also scan suspicious calls to these API functions. After that, when detected, the program sends a message about a suspicious call and requests permission. In case of failure, Mimikatz is terminated.

When considering the injection method into the LSASS process, it is necessary to consider the implementation. There are certain problems, since searching directly for the Mimikatz process is not the best idea. It is necessary to make a monitoring system that will monitor the system in real time for connecting applications and signal if suspicious software is connected. At the moment, monitoring is configured so that it compares the newly launched application with a list of trusted applications, which are stored in a separate array, where the full paths to the executable files of trusted applications are prescribed in order to avoid substitution of applications by name. An example of how monitoring works is shown in the figure 11.

```
msedge.exe
msedge.exe
msedge.exe
msedge.exe
msedge.exe
identity_helper.exe
msedge.exe
msedge.exe
msedge.exe
msedge.exe
backgroundTaskHost.exe
msedge.exe
identity_helper.exe
msedge.exe
msedge.exe
msedge.exe
msedge.exe
RuntimeBroker.exe
apimonitor-x64.exe
dllhost.exe
mimikatz.exe
WARNING
conhost.exe
dllhost.exe
```

Figure 11 — monitoring

The system opposes directly at the stage of extracting user data, and interaction does not occur too early at the initial stage, where rights to work in debugging mode are requested, the call and launch of the application itself is not blocked, since tracking of each application is impossible, since there are several different applications, moreover, Mimikatz itself is easy to rebuild under a different name and with other key functions.

However, the interaction does not occur too late, when some of the information has already been extracted, since it is necessary to take into account the delay of the system in responding to the implementation. For example, if interact with very specific functions that can only be called and used by malicious software, it may not cover the protected information sufficiently, which is why the system will not have time to react in time and redirect the malicious software to the safe side, preventing it from revealing the user's credentials.

Interaction with the LsaOpenPolicy() function is the most optimal in this case, since this function belongs to a user-level library, unlike deeper kernel-level libraries, and it is much more convenient to interact with it

The third way to prevent mimitatz-like attack is *ALPC blocking* — blocking connections between SamConnect() and LSASS. For this method we need to do this:

1. change the field Token.Privileges.Enabled

2. change the field Token.EnabledByDefault

3. change the field IntegrityLevelIndex

So, mimikatz can't connect to LSASS without ALPC connection and can't steal password information.

But this method requires extreme accuracy. If you do not return the settings to their original state before logging out, that is, you do not return the LSASS settings to their previous state, then a full reboot may be required.

## 5 CONCLUSION

In conclusion, we want to highlight the following:

- The Windows operating system is actively fighting the dumping of credentials and offers its own solutions that, in certain cases, help prevent attacks.

- There are methods that can bypass Windows protection systems, so continuous improvement of data dumping protection techniques is required.

- When protecting data, an integrated approach is necessary, since successful attacks on LSASS always occur not only using Mimikatz, but also other various techniques.

In the future, it is planned to test and compare all the methods mentioned above, primarily method 2 with injection into the LSASS process in order to identify suspicious function calls samsrv.dll and samlib.dll.